\documentclass[eqsecnum,superscriptaddress,11pt]{revtex4}
\pdfoutput=1
\usepackage{amsmath}
   \usepackage{bm}
\usepackage{amsthm,amscd}
\usepackage{mathrsfs}
\usepackage{verbatim}
\usepackage{amsfonts,amsmath,latexsym,amssymb,txfonts,bm}
\usepackage[american]{babel}
\usepackage{dsfont}
\usepackage{graphicx}
\usepackage{latexsym}
\newcommand{\diff}{\mathrm{d}}

\def\beq{\begin{eqnarray}}
\def\eeq{\end{eqnarray}}

\begin{document}
\title{Casimir Pistons with General Boundary Conditions}

\author{Guglielmo Fucci\footnote{Electronic address: fuccig@ecu.edu}}
\affiliation{Department of Mathematics, East Carolina University, Greenville, NC 27858 USA}



\date{\today}
\vspace{2cm}
\begin{abstract}

In this work we analyze the Casimir energy and force for a scalar field endowed with general self-adjoint boundary
conditions propagating in a higher dimensional piston configuration. The piston is constructed as a direct product $I\times N$,
with $I=[0,L]\subset\mathbb{R}$ and $N$ a smooth, compact Riemannian manifold with or without boundary.
The study of the Casimir energy and force for this configuration is performed by employing the spectral zeta function regularization technique.
The obtained analytic results depend explicitly on the spectral zeta function associated with the manifold $N$ and the
parameters describing the general boundary conditions imposed. These results are then specialized to the case in which the manifold $N$
is a $d$-dimensional sphere.

\end{abstract}
\maketitle

\section{Introduction}

The Casimir effect refers to a broad set of phenomena which are caused by changes to the vacuum energy of a quantum field due to the
presence of either boundaries or non-dynamical external fields. The effect was first predicted by Casimir in \cite{casimir48} who analyzed
a configuration consisting of two parallel plates. However, in general the Casimir effect manifests itself through the appearance of a net force between
two neutral objects \cite{bordag09}. Due to its theoretical as well as experimental importance the Casimir effect has been a subject of quite intense
research for the past several decades (see, e.g., \cite{bordag01,bordag09,casimir48,milton01,plunien86} and references therein).
For the vast majority of configurations, calculations of the Casimir energy lead, unfortunately, to meaningless divergent quantities that require
regularization and subsequent renormalization \cite{blau88,bytsenko03,elizalde94,elizalde}. This can be accomplished through a number of techniques \cite{bordag09} which constitute rather standard tools in quantum field theory.

Piston configurations were first introduced in the well-known work of Cavalcanti \cite{caval04} who analyzed the
Casimir effect for a massless scalar field propagating in a rectangular box divided in two regions
by a movable piston. He found that the Casimir energy generates a force that tends to move the piston to the closest wall.
Since his seminal work, piston configurations have attracted widespread interest and have been the topic of a substantial number of publications.
One of the main reasons that make piston configurations such an interesting subject of study lies in the fact that although their Casimir energy might be divergent,
the corresponding Casimir force acting on the piston is well defined and free of divergences. This, however,
becomes no longer generally true when one considers piston configurations with non-vanishing curvature \cite{fucci11,fucci11b,fucci12}.

Several types of geometric configurations for Casimir pistons have been considered throughout the literature, see e.g. \cite{barton06,edery06,hertz07,li97,marachevsky07,kirsten09}.
Almost all of them, though, analyze the Casimir effect for quantum fields subject to standard boundary conditions.
Here, by standard boundary conditions we intend Dirichlet, Neumann, and Robin boundary conditions.
Other sets of boundary conditions that have also been considered are hybrid, or mixed, boundary conditions
which are obtained by imposing different standard boundary conditions on different sides of the two chambers of the piston.
The main purpose for studying such a wide variety of piston configurations endowed with standard boundary conditions
is to analyze the dependence of the Casimir energy and force on the particular geometry of the system and
on the boundary conditions imposed. Lately, repulsive Casimir forces have become a subject of major interest due
to their relevance in the development of microelectromechanical devices (MEMS). As a result of their microscopic size such devices
are afflicted by the problem of stiction, in which different components of the device adhere to each other due to an attractive
Casimir force. It is, therefore, of particular importance to understand what type of boundary conditions, which are used
to model properties of materials, need to be imposed in order to obtain a repulsive, or even vanishing, Casimir force.
Configurations that lead to a repulsive Casimir force have been analyzed, for instance, in
\cite{caruso91,elizalde09,milton12,munoz13,piro09,schmidt08}.

In this work we consider a massless scalar field propagating in a higher dimensional piston configuration endowed with general self-adjoint
boundary conditions. Clearly, the general boundary conditions considered here contain, as particular cases, the standard
and hybrid boundary conditions mentioned earlier. By using the spectral zeta function regularization technique
we compute the Casimir energy of the piston configuration and the ensuing Casimir force on the piston itself.
The expression we obtain for the Casimir force depends explicitly on the general boundary conditions imposed which, in turn,
are described by six independent parameters. Our results are therefore suitable for analyzing how the Casimir force changes,
both in magnitude and sign, when any of the parameters describing the general boundary conditions vary.
The results obtained in this work can be used to determine a range of values for the parameters in the boundary conditions
that results on a repulsive force of the piston from one or both ends of the piston configuration.
These values provide a set of particular boundary conditions which could be utilized for selecting specific
materials in the design and development of microelectromechanical devices. It is important to mention that the
study of the Casimir energy for massless scalar fields endowed with general boundary conditions have been conducted, for instance, in \cite{asorey13}.
Their analysis focuses on homogeneous parallel plates embedded in $\mathbb{R}^{D}$. For this configuration the variation
of the Casimir force on the plates with respect to the general boundary conditions is explicitly shown.
Our work, instead, considers higher dimensional piston configurations with general boundary conditions
and, hence, extends the results in \cite{asorey13} to Casimir pistons.

The outline of the paper is as follows. In the next Section we describe in details the piston configuration with general boundary conditions
and represent the associated spectral zeta function in terms of a complex integral. In Section \ref{sec3} the analytic
continuation of the spectral zeta function to a region containing
the point $s=-1/2$ is developed. Section \ref{sec4} focuses on the necessary modifications to the spectral zeta function arising from the
presence of zero modes on the manifold $N$. In the subsequent two Sections the Casimir energy and force on the piston is obtained and numerical
results for the force are provided for some specific examples. The conclusions point to the main results and outline possible additional studies along the lines developed
in this work.

\section{Casimir Piston and the Zeta Function}

We consider a bounded, $D$-dimensional manifold $M$ of the type $M=I\times N$, with $I=[0,L]\subset\mathbb{R}$ a closed
interval of the real line and $N$ a smooth, compact, $d$-dimensional Riemannian manifold with or without boundary.
Obviously, the above remarks imply that $D=d+1$. The manifold $M$ can be used to
construct a piston configuration as follows (see also \cite{fucci12}): For all points $a\in I$ we define the manifold $N_{a}$
to be the cross-section of $M$ positioned at $a$. The manifold $M$ can then be divided, along $N_{a}$, in two regions, or chambers,
denoted by $M_{I}$ and $M_{II}$. Both $M_{I}$ and $M_{II}$ are $D$-dimensional compact Riemannian manifolds having the geometry
of a product, namely $M_{I}=[0,a]\times N$ and $M_{II}=(a,L]\times N$. Since $\partial M_{I}=N_{0} \cup N_{a}\cup ([0,a]\times\partial N)$
and $\partial M_{II}=N_{a} \cup N_{L}\cup ((a,L]\times\partial N)$,
the two chambers $M_{I}$ and $M_{II}$ have the cross-section $N_{a}$ as a common boundary. The manifold $M=M_{I}\cup M_{II}$ represents the piston
configuration with $N_{a}$ describing the piston itself.

The dynamics of a massless scalar field $\phi$ propagating on the piston $M$ is described by the differential equation
\begin{equation}\label{1}
-\Delta_{M}\phi=\alpha^{2}\phi\;,
\end{equation}
with $\Delta_{M}$ being the Laplace operator acting on square-integrable scalar functions $\phi\in L^{2}(M)$. In the appropriate
system of coordinates, equation (\ref{1}) can be explicitly written as
\begin{equation}\label{2}
-\left(\frac{\diff^{2}}{\diff x^{2}}+\Delta_{N}\right)\phi=\alpha^{2}\phi\;,\quad
\end{equation}
where we have denoted by $\Delta_{N}$ the Laplacian on the manifold $N$. In the framework of the Casimir effect,
the fields propagating in one region of the piston are independent from the fields propagating in the
other region. This implies that the differential equation (\ref{2}) has to be solved in region $I$ and region $II$
separately. To this end, we denote with $\alpha_{I}$ the eigenvalues of (\ref{2}) in region $I$ and with
$\alpha_{II}$ the eigenvalues of the same equation in region $II$, with $\phi_{I}$ and $\phi_{II}$ being the corresponding
eigenfunctions. The solutions $\phi_{i}$, where $i=\{I,II\}$, can be written as a product $\phi_{i}=f_{i}(x,\nu)\Phi(X)$, where
$\Phi(X)$ represent the eigenfunctions of the Laplacian on $N$, namely
\begin{equation}\label{3}
-\Delta_{N}\Phi(X)=\nu^{2}\Phi(X)\;.
\end{equation}
In this case, the functions $f_{i}(\nu,x)$ satisfy the following ordinary differential equation
\begin{equation}\label{4}
\left(-\frac{\diff^{2}}{\diff x^{2}}+\nu^{2}-\alpha_{i}^{2}\right)f_{i}(\nu,x)=0\;,
\end{equation}
in each region.

By using the eigenvalues $\alpha_{i}^{2}$ we define the spectral zeta function $\zeta_{i}(s)$ associated with either region $I$ or $II$
as
\begin{equation}\label{5}
\zeta_{i}(s)=\sum_{n=1}^{\infty}\alpha_{i}^{-2s}\;,
\end{equation}
valid in the half-plane $\Re(s)>D/2$.
The spectral zeta function associated with the piston configuration is then expressed as the sum
\begin{equation}\label{6}
\zeta(s)=\zeta_{I}(s)+\zeta_{II}(s)\;,
\end{equation}
of the zeta functions in region $I$ and $II$. Once the spectral zeta function (\ref{6}) has been analytically
continued to a region containing the point $s=-1/2$, it can be used to find the Casimir energy $E_{\textrm{Cas}}$ through the formula \cite{bordag01,bordag09,bytsenko03,elizalde94,elizalde,kirsten01,milton01}
\begin{equation}\label{7}
E_{\textrm{Cas}}=\lim_{\varepsilon\to 0}\frac{\mu^{2\varepsilon}}{2}\zeta\left(\varepsilon-\frac{1}{2}\right)\;.
\end{equation}
Since the spectral zeta function generally develops a simple pole at the point $s=-1/2$ \cite{kirsten01}, the Casimir energy
can be rewritten as
\begin{equation}\label{8}
E_{\textrm{Cas}}=\frac{1}{2}\textrm{FP}\,\zeta\left(-\frac{1}{2}\right)+\frac{1}{2}\left(\frac{1}{\varepsilon}+\ln\mu^{2}\right)\textrm{Res}\,\zeta\left(-\frac{1}{2}\right)+O(\varepsilon)\;,
\end{equation}
with $\textrm{FP}$ and $\textrm{Res}$ denoting, respectively, the finite part and the residue. The Casimir energy is, hence, well defined
when the residue of the spectral zeta function at $s=-1/2$ vanishes. In the case of piston configurations, the Casimir energy depends explicitly
on the position of the piston $a$. The Casimir force acting on the piston can then be computed according to the formula
\begin{equation}\label{9}
F_{\textrm{Cas}}(a)=-\frac{\partial}{\partial a}E_{\textrm{Cas}}(a)\;.
\end{equation}
It is clear that the Casimir force on the piston is well defined only when the residue of the zeta function in (\ref{8})
is independent of the position of the piston $a$.

In order to analyze the spectral zeta functions $\zeta_{I}(s)$ and $\zeta_{II}(s)$ we need the eigenvalues $\alpha_{I}$ and $\alpha_{II}$.
These eigenvalues are not explicitly known in general, but implicit equations for them can be found by imposing appropriate boundary conditions
to the differential equation (\ref{4}) in region $I$ and region $II$. In this work we will consider the most general separated boundary conditions
that, once imposed to (\ref{4}), lead to a self-adjoint boundary value problem. This boundary conditions can be expressed in region $I$ as \cite{zettl}
\begin{eqnarray}\label{10}
A_{1}f_{I}(\nu,0)+A_{2}f^{\prime}_{I}(\nu,0)&=&0\;,\nonumber\\
B_{1}f_{I}(\nu,a)-B_{2}f^{\prime}_{I}(\nu,a)&=&0\;,
\end{eqnarray}
and in region $II$ as
\begin{eqnarray}\label{11}
B_{1}f_{II}(\nu,a)-B_{2}f^{\prime}_{II}(\nu,a)&=&0\;,\nonumber\\
C_{1}f_{II}(\nu,L)+C_{2}f^{\prime}_{II}(\nu,L)&=&0\;,
\end{eqnarray}
where $A_{1},A_{2},B_{1},B_{2},C_{1},C_{2}\in\mathbb{R}$ with the conditions $(A_{1},A_{2})\neq(0,0)$, $(B_{1},B_{2})\neq (0,0)$, and $(C_{1},C_{2})\neq (0,0)$.
We can, now, analyze the solutions of the differential equation (\ref{4}) endowed with the boundary conditions (\ref{10}) in region $I$ and (\ref{11}) in region $II$.
The general solution of (\ref{4}) is
\begin{equation}\label{12}
f_{i}(\nu,x)=a\,e^{i \sqrt{\alpha_{i}^{2}-\nu^{2}}x}+b\,e^{i \sqrt{\alpha_{i}^{2}-\nu^{2}}x}\;,
\end{equation}
with the coefficients $a$ and $b$ to be determined by imposing the boundary conditions.

In region $I$ we choose a solution $f_{I}(\nu,x)$ that
satisfies the following condition at $x=0$:
\begin{equation}\label{13}
f_{I}(\nu,0)=-A_{2}\;,\quad f_{I}^{\prime}(\nu,0)=A_{1}\;.
\end{equation}
It is not difficult to realize that a solution $f_{I}(\nu,x)$ satisfying (\ref{13}) also automatically satisfies the first equation
of the boundary conditions (\ref{10}). Applying (\ref{13}) to the general solution (\ref{12}) gives the following result
\begin{equation}\label{14}
f_{I}(\nu,x)=\frac{A_{1}}{\sqrt{\alpha_{I}^{2}-\nu^{2}}}\sin\left(\sqrt{\alpha_{I}^{2}-\nu^{2}}\, x\right)-A_{2}\cos\left(\sqrt{\alpha_{I}^{2}-\nu^{2}}\, x\right)\;.
\end{equation}
By imposing now the boundary condition at $x=a$ to the solution (\ref{14}) we obtain an implicit equation
for the eigenvalues $\alpha_{I}$ in region $I$
\begin{equation}\label{15}
\Omega^{I}_{\nu}(\alpha,a)=\left(\frac{A_{1}B_{1}}{\sqrt{\alpha_{I}^{2}-\nu^{2}}}-A_{2}B_{2}\sqrt{\alpha_{I}^{2}-\nu^{2}}\right)\sin\left(\sqrt{\alpha_{I}^{2}-\nu^{2}}\, a\right)
-\left(A_{2}B_{1}+A_{1}B_{2}\right)\cos\left(\sqrt{\alpha_{I}^{2}-\nu^{2}}\, a\right)=0\;.
\end{equation}

In region $II$ we consider a solution, denoted by $f_{II}(\nu,x)$, which satisfies the conditions
\begin{equation}\label{16}
f_{II}(\nu,a)=B_{2}\;,\quad f^{\prime}_{II}(\nu,a)=B_{1}\;.
\end{equation}
Once again, a solution that satisfies (\ref{16}) also satisfies the first condition in (\ref{11}) and has the form
\begin{equation}\label{17}
f_{II}(\nu,x)=\frac{B_{1}}{\sqrt{\alpha_{II}^{2}-\nu^{2}}}\sin\left[\sqrt{\alpha_{II}^{2}-\nu^{2}}\, (x-a)\right]+B_{2}\cos\left[\sqrt{\alpha_{II}^{2}-\nu^{2}}\, (x-a)\right]\;.
\end{equation}
Imposing the second boundary condition in (\ref{11}) leads to the following implicit equation for the eigenvalues
$\alpha_{II}$ in region $II$
\begin{eqnarray}\label{18}
\Omega^{II}_{\nu}(\alpha,a)&=&\left(\frac{B_{1}C_{1}}{\sqrt{\alpha_{II}^{2}-\nu^{2}}}-B_{2}C_{2}\sqrt{\alpha_{II}^{2}-\nu^{2}}\right)\sin\left[\sqrt{\alpha_{II}^{2}-\nu^{2}}\, (L-a)\right]\nonumber\\
&+&\left(B_{2}C_{1}+B_{1}C_{2}\right)\cos\left[\sqrt{\alpha_{II}^{2}-\nu^{2}}\, (L-a)\right]=0\;.
\end{eqnarray}

The solutions of (\ref{15}) and (\ref{18}) are simple and either real or purely imaginary \cite{romeo02,teo09}.
Since here we consider eigenvalues of a self-adjoint boundary value problem, we restrict our analysis to the case in which
all zeroes of (\ref{15}) and (\ref{18}) are real. A discussion of the case in which purely imaginary zeroes are present can be
found in \cite{romeo02}. The purely imaginary zeroes of $\Omega^{I}_{\nu}(\alpha,a)$ correspond to the real zeroes of
$\Omega^{I}_{\nu}(i\alpha,a)$, namely
\begin{equation}\label{19}
\Omega^{I}_{\nu}(i\alpha,a)=\left(\frac{A_{1}B_{1}}{\sqrt{\alpha_{I}^{2}+\nu^{2}}}+A_{2}B_{2}\sqrt{\alpha_{I}^{2}+\nu^{2}}\right)\sinh\left(\sqrt{\alpha_{I}^{2}+\nu^{2}}\, a\right)
-\left(A_{2}B_{1}+A_{1}B_{2}\right)\cosh\left(\sqrt{\alpha_{I}^{2}+\nu^{2}}\, a\right)\;.
\end{equation}
The real zeroes of (\ref{19}) can be found as solution of the equation
\begin{equation}\label{20}
\frac{\tanh\omega}{\omega}=\frac{A_{2}B_{1}+A_{1}B_{2}}{a A_{1}B_{1}\left(1+\frac{A_{2}B_{2}}{a^{2}A_{1}B_{1}}\omega^{2}\right)}\;,
\end{equation}
where we have set, for typographical convenience, $\omega=\sqrt{\alpha_{I}^{2}+\nu^{2}}\, a$. From (\ref{20}) we can conclude that
$\Omega^{I}_{\nu}(i\alpha,a)$ has no real zeroes, and hence $\Omega^{I}_{\nu}(\alpha,a)$ in (\ref{15}) has no purely
imaginary zeroes, if
\begin{equation}\label{21}
\left\{\frac{A_{2}B_{2}}{a^{2}A_{1}B_{1}}\leq 0\;,\frac{1}{a}\left(\frac{A_{2}}{A_{1}}+\frac{B_{2}}{B_{1}}\right)\geq 1\right\}\;,\quad \textrm{or}\quad
\left\{\frac{A_{2}}{a A_{1}}\leq 0\;, \frac{B_{2}}{a B_{1}}\leq 0\right\}\;, \quad A_{1}B_{1}\neq0\;,
\end{equation}
and
\begin{equation}\label{21a}
\frac{B_{2}}{a B_{1}}<0\;,\quad A_{1}=0\;,\quad\textrm{and}\quad \frac{A_{2}}{a A_{1}}<0\;,\quad B_{1}=0\;.
\end{equation}
Under the above conditions we represent
the spectral zeta function $\zeta_{I}(s,a)$ in terms of a contour integral in the complex plane valid for $\Re(s)>D/2$ as \cite{bordag96a,bordag96b,kirsten01}
\begin{equation}\label{22}
\zeta_{I}(s,a)=\frac{1}{2\pi i}\sum_{\nu}d(\nu)\int_{\gamma_{I}}\kappa^{-2s}\frac{\partial}{\partial \kappa}\ln\Omega^{I}_{\nu}(\kappa,a)\diff\kappa\;,
\end{equation}
where $d(\nu)$ denotes the multiplicity of the eigenvalues $\nu$ of the Laplacian on the manifold $N$, and $\gamma$ is a contour enclosing in the counterclockwise direction
all the real zeroes of $\Omega^{I}_{\nu}(\kappa,a)$.
By performing the change of variables $\kappa=z\nu$ and by deforming the contour $\gamma_{I}$ to the imaginary axis
we obtain \cite{kirsten01}
\begin{equation}\label{23}
\zeta_{I}(s,a)=\sum_{\nu}d(\nu)\zeta^{\nu}_{I}(s,a)\;,
\end{equation}
where
\begin{equation}\label{24}
\zeta^{\nu}_{I}(s,a)=\frac{\sin\pi s}{\pi}\nu^{-2s}\int_{0}^{\infty}z^{-2s}\frac{\partial}{\partial z}\ln\Omega^{I}_{\nu}(i\nu z,a)\diff z\;.
\end{equation}
The integral representation (\ref{24}) is valid only in a vertical strip of the complex plane. The region of convergence of the integral (\ref{24})
can be found by analyzing the behavior of the integral as $z\to \infty$ and as $z\to 0$. The behaviour of the function
$\Omega^{I}_{\nu}(i\nu z,a)$ for large values of the variable $z$ is rendered manifest by rewriting its expression, by using (\ref{19}), as
\begin{equation}\label{25}
\Omega^{I}_{\nu}(i\nu z,a)=e^{\nu\sqrt{1+z^{2}}\,a} \left(\frac{A_{1}B_{1}}{2\nu\sqrt{1+z^{2}}}-\frac{A_{1}B_{2}+A_{2}B_{1}}{2}+\frac{A_{2}B_{2}}{2}\nu\sqrt{1+z^{2}}\right)
\left[1+\epsilon(\nu,z,a)\right]\;,
\end{equation}
where $\epsilon(\nu,z,a)$ represent exponentially small terms. From (\ref{25}) it is not difficult to prove that,
as $z\to\infty$, we have the following behavior
\begin{equation}\label{26}
z^{-2s}\frac{\partial}{\partial z}\ln\Omega^{I}_{\nu}(i\nu z,a)\sim \nu a z^{-2s}\;.
\end{equation}
This implies that the integral (\ref{24}) converges at the upper limit of integration when $\Re(s)>1/2$. As $z\to 0$ we have, instead,
\begin{eqnarray}\label{27}
\Omega^{I}_{\nu}(i\nu z,a)&=&\left(\frac{A_{1}B_{1}}{\nu}+A_{2}B_{2}\nu\right)\sinh\nu a
-\left(A_{2}B_{1}+A_{1}B_{2}\right)\cosh\nu a\nonumber\\
&+&\frac{z^{2}}{2\nu}\left[\nu a(A_{1}B_{1}+\nu^{2}A_{2}B_{2})\cosh\nu a-\left(A_{1}B_{1}-\nu^{2}A_{2}B_{2}+a\nu^{2}(A_{1}B_{2}+A_{2}B_{1})\right)\right]\sinh\nu a
+O(z^{4})\;.\nonumber\\
\end{eqnarray}
Under the conditions (\ref{21}) and (\ref{21a}), the first term of the expansion of $\Omega^{I}_{\nu}(i\nu z,a)$ in (\ref{27})
is non-vanishing and, therefore, the behavior of the integrand in (\ref{24}) is
\begin{equation}\label{28}
z^{-2s}\frac{\partial}{\partial z}\ln\Omega^{I}_{\nu}(i\nu z,a)\sim z^{-2s+1}\;,
\end{equation}
which implies that the integral (\ref{24}) converges at the lower limit of integration for $\Re(s)<1$. The above remarks allow us to conclude that
the integral representation of $\zeta^{\nu}_{I}(s,a)$ in (\ref{24}) is valid in the region $1/2<\Re(s)<1$ of the complex plane.

The procedure developed above for analyzing the spectral zeta function in region $I$ can be repeated for region $II$. In fact, equation
(\ref{18}) has only real zeroes if the following function has no real zeroes
\begin{eqnarray}\label{29}
\Omega^{II}_{\nu}(i\alpha,a)&=&\left(\frac{B_{1}C_{1}}{\sqrt{\alpha_{II}^{2}+\nu^{2}}}+B_{2}C_{2}\sqrt{\alpha_{II}^{2}+\nu^{2}}\right)\sinh\left[\sqrt{\alpha_{II}^{2}+\nu^{2}}\, (L-a)\right]\nonumber\\
&+&\left(B_{2}C_{1}+B_{1}C_{2}\right)\cosh\left[\sqrt{\alpha_{II}^{2}+\nu^{2}}\, (L-a)\right]\;.
\end{eqnarray}
This is the case if the conditions
\begin{equation}\label{30}
\left\{\frac{B_{2}C_{2}}{(L-a)^{2}B_{1}C_{1}}\leq 0\;,\frac{1}{L-a}\left(\frac{B_{2}}{ B_{1}}+\frac{C_{2}}{C_{1}}\right)\leq -1\right\}\;,\quad \textrm{or}\quad
\left\{\frac{B_{2}}{(L-a) B_{1}}\geq 0\;, \frac{C_{2}}{(L-a) C_{1}}\geq 0\right\}\;, \quad B_{1}C_{1}\neq0\;,
\end{equation}
and
\begin{equation}\label{31}
\frac{B_{2}}{(L-a) B_{1}}>0\;,\quad C_{1}=0\;,\quad\textrm{and}\quad \frac{C_{2}}{(L-a) C_{1}}>0\;,\quad B_{1}=0\;,
\end{equation}
are satisfied.
Under these conditions, the spectral zeta function $\zeta_{II}(s,a)$ can be represented as a contour integral similar to the one in (\ref{22}).
After deforming the contour to the imaginary axis we obtain the expression
\begin{equation}\label{32}
\zeta_{II}(s,a)=\sum_{\nu}d(\nu)\zeta^{\nu}_{II}(s,a)\;,
\end{equation}
with
\begin{equation}\label{33}
\zeta^{\nu}_{II}(s,a)=\frac{\sin\pi s}{\pi}\nu^{-2s}\int_{0}^{\infty}z^{-2s}\frac{\partial}{\partial z}\ln\Omega^{II}_{\nu}(i\nu z,a)\diff z\;,
\end{equation}
which, by using the same argument outlined before, can be shown to be valid in the strip $1/2<\Re(s)<1$. According to the definition (\ref{7}),
in order to compute the Casimir energy, and the corresponding force on the piston, one needs the spectral zeta function in a
neighborhood of $s=-1/2$. Since this point does not belong to the region where the integral representations (\ref{24}) and (\ref{33}) are valid,
we have to analytically continue $\zeta_{I}(s,a)$ and $\zeta_{II}(s,a)$ to the region $\Re(s)\leq 1/2$.

\section{Analytic Continuation of the zeta function}\label{sec3}

The analytic continuation of the functions $\zeta_{I}(s,a)$ and $\zeta_{II}(s,a)$ to a region to the left of $\Re(s)=1/2$
is obtained through the addition and subtraction of a suitable number of terms of the asymptotic expansion of $\ln\Omega^{I}_{\nu}(i\nu z,a)$,
respectively $\ln\Omega^{II}_{\nu}(i\nu z,a)$, as $\nu\to\infty$ with $z=\kappa/\nu$ fixed \cite{kirsten01}. To construct the desired
asymptotic expansion we utilize the expression (\ref{25}) in region $I$ to obtain
\begin{equation}\label{34}
\ln\Omega^{I}_{\nu}(i\nu z,a)=\nu\sqrt{1+z^{2}}a+\ln\left[\frac{A_{1}B_{1}}{2\nu\sqrt{1+z^{2}}}-\frac{A_{1}B_{2}+A_{2}B_{1}}{2}+\frac{A_{2}B_{2}}{2}\nu\sqrt{1+z^{2}}\right]
+\ln\left[1+\epsilon_{I}(\nu,z,a)\right]\;,
\end{equation}
which, by introducing the function,
\begin{equation}
\delta(x)=\left\{\begin{array}{ll}
1 & \textrm{if}\; x=0\\
0 & \textrm{if}\; x\neq 0
\end{array}\right.\;,
\end{equation}
can be rewritten in a form suitable for a large-$\nu$ expansion uniform in the variable $z$
\begin{eqnarray}\label{35}
\ln\Omega^{I}_{\nu}(i\nu z,a)&=&\nu\sqrt{1+z^{2}}a+\left[1-\delta(A_{2}B_{2})-\delta(A_{2})\delta(B_{2})\right]\ln\left(\nu\sqrt{1+z^{2}}\right)
+\left[1-\delta(A_{2}B_{2})\right]\ln\left(\frac{A_{2}B_{2}}{2}\right)\nonumber\\
&+&\delta(A_{2})\delta(B_{2})\ln\left(\frac{A_{1}B_{1}}{2}\right)+\left[\delta(A_{2}B_{2})-\delta(A_{2})\delta(B_{2})\right]\ln\left(-\frac{A_{1}B_{2}\delta(A_{2})+A_{2}B_{1}\delta(B_{2})}{2}\right)\nonumber\\
&+&\left[1-\delta(A_{2}B_{2})\right]\ln\left[1-\frac{A_{1}B_{2}+A_{2}B_{1}}{A_{2}B_{2}}\frac{1}{\nu\sqrt{1+z^{2}}}+\frac{A_{1}B_{1}}{A_{2}B_{2}}\frac{1}{\nu^{2}(1+z^{2})}\right]\nonumber\\
&+&\left[\delta(A_{2}B_{2})-\delta(A_{2})\delta(B_{2})\right]\ln\left[1-\frac{A_{1}B_{1}}{A_{1}B_{2}\delta(A_{2})+A_{2}B_{1}\delta(B_{2})}\frac{1}{\nu\sqrt{1+z^{2}}}\right]
+\ln\left[1+\epsilon_{I}(\nu,z,a)\right]\;.\nonumber\\
\end{eqnarray}

By using the small-$x$ asymptotic expansion of $\ln(1+x)$ we obtain the following large-$\nu$ expansion for the last term in (\ref{35}):
\begin{equation}\label{36}
\ln\left[1-\frac{A_{1}B_{1}}{A_{1}B_{2}\delta(A_{2})+A_{2}B_{1}\delta(B_{2})}\frac{1}{\nu\sqrt{1+z^{2}}}\right]\sim -\sum_{n=1}^{\infty}\frac{D_{n}}{n}\frac{1}{\nu^{n}(1+z^{2})^{\frac{n}{2}}}\;,
\end{equation}
with the coefficients $D_{n}$ defined as
\begin{equation}\label{37}
D_{n}=\left(\frac{A_{1}B_{1}}{A_{1}B_{2}\delta(A_{2})+A_{2}B_{1}\delta(B_{2})}\right)^{n}\;.
\end{equation}
The large-$\nu$ asymptotic expansion of the second to the last term in (\ref{35}) can be obtained by using the small-$x$ expansion
\begin{equation}\label{38}
\ln\left(1+Ax+Bx^{2}\right)\sim \sum_{l=1}^{\infty}C_{l}x^{l}\;,
\end{equation}
where $A$ and $B$ are real constants and
\begin{equation}\label{39}
C_{l}=(-1)^{l-1}\sum_{q=0}^{\left[\frac{l}{2}\right]}\frac{(-1)^{q}}{l-q}\binom{l-q}{q}A^{l-2q}B^{q}\;,
\end{equation}
with $[x]$ denoting the integer part of $x$. More precisely we have
\begin{equation}\label{40}
\ln\left[1-\frac{A_{1}B_{2}+A_{2}B_{1}}{A_{2}B_{2}}\frac{1}{\nu\sqrt{1+z^{2}}}+\frac{A_{1}B_{1}}{A_{2}B_{2}}\frac{1}{\nu^{2}(1+z^{2})}\right]\sim
\sum_{l=1}^{\infty}\frac{E_{l}}{\nu^{l}(1+z^{2})^{\frac{l}{2}}}\;,
\end{equation}
where the expression for $E_{l}$ can be obtained from (\ref{39}). In fact,
\begin{eqnarray}\label{41}
E_{l}&=&-\sum_{q=0}^{\left[\frac{l}{2}\right]}\frac{(-1)^{q}}{l-q}\binom{l-q}{q}\left(\frac{A_{1}B_{1}}{A_{2}B_{2}}\right)^{q}\left(\frac{A_{1}}{A_{2}}+\frac{B_{1}}{B_{2}}\right)^{l-2q}\nonumber\\
&=&-\frac{1}{l}\left(\frac{A_{1}}{A_{2}}+\frac{B_{1}}{B_{2}}\right)^{l}-\sum_{q=1}^{\left[\frac{l}{2}\right]}\sum_{j=0}^{l-1}\frac{(-1)^{q}}{l-q}\binom{l-q}{q}\binom{l-2q}{j}
\left(\frac{A_{1}}{A_{2}}\right)^{l-q-j}\left(\frac{B_{1}}{B_{2}}\right)^{q+j}\;.
\end{eqnarray}
By setting $k=q+j$ the double sum appearing above can be written in terms of one to give the following simple expression for $E_{l}$:
\begin{equation}\label{42}
E_{l}=-\frac{1}{l}\left(\frac{A_{1}}{A_{2}}+\frac{B_{1}}{B_{2}}\right)^{l}+\sum_{k=1}^{l-1}\frac{1}{l}\binom{l}{k}\left(\frac{A_{1}}{A_{2}}\right)^{k}\left(\frac{B_{1}}{B_{2}}\right)^{l-k}=-\frac{1}{l}\left[\left(\frac{A_{1}}{A_{2}}\right)^{l}+\left(\frac{B_{1}}{B_{2}}\right)^{l}\right]\;.
\end{equation}

The large-$\nu$ expansion of (\ref{35}) uniform in $z$ can now be written as
\begin{eqnarray}\label{43}
\ln\Omega^{I}_{\nu}(i\nu z,a)&\sim&\nu\sqrt{1+z^{2}}a+\left[1-\delta(A_{2}B_{2})-\delta(A_{2})\delta(B_{2})\right]\ln\left(\nu\sqrt{1+z^{2}}\right)
+\left[1-\delta(A_{2}B_{2})\right]\ln\left(\frac{A_{2}B_{2}}{2}\right)\nonumber\\
&+&\delta(A_{2})\delta(B_{2})\ln\left(\frac{A_{1}B_{1}}{2}\right)+\left[\delta(A_{2}B_{2})-\delta(A_{2})\delta(B_{2})\right]\ln\left(-\frac{A_{1}B_{2}\delta(A_{2})+A_{2}B_{1}\delta(B_{2})}{2}\right)\nonumber\\
&+&\sum_{k=1}^{\infty}\frac{F_{k}}{\nu^{k}(1+z^{2})^{\frac{k}{2}}}\;,
\end{eqnarray}
where
\begin{equation}\label{43a}
F_{k}=\left[1-\delta(A_{2}B_{2})\right]E_{k}
+\left[\delta(A_{2}B_{2})-\delta(A_{2})\delta(B_{2})\right]\frac{D_{k}}{k}\;,
\end{equation}
and we have discarded exponentially decreasing terms.

In region $II$ we consider the function
\begin{equation}\label{44}
\ln\Omega^{II}_{\nu}(i\nu z,a)=\nu\sqrt{1+z^{2}}(L-a)+\ln\left[\frac{B_{1}C_{1}}{2\nu\sqrt{1+z^{2}}}+\frac{B_{1}C_{2}+B_{2}C_{1}}{2}+\frac{B_{2}C_{2}}{2}\nu\sqrt{1+z^{2}}\right]
+\ln\left[1+\epsilon_{II}(\nu,z,a)\right]\;.
\end{equation}
By utilizing the same procedure that led to the large-$\nu$ asymptotic expansion of $\ln\Omega^{I}_{\nu}(i\nu z,a)$ in (\ref{43}) we obtain
for $\ln\Omega^{II}_{\nu}(i\nu z,a)$ the asymptotic expansion valid for large $\nu$ and fixed $z$
\begin{eqnarray}\label{45}
\ln\Omega^{II}_{\nu}(i\nu z,a)&\sim&\nu\sqrt{1+z^{2}}(L-a)+\left[1-\delta(B_{2}C_{2})-\delta(B_{2})\delta(C_{2})\right]\ln\left(\nu\sqrt{1+z^{2}}\right)
+\left[1-\delta(B_{2}C_{2})\right]\ln\left(\frac{B_{2}C_{2}}{2}\right)\nonumber\\
&+&\delta(B_{2})\delta(C_{2})\ln\left(\frac{B_{1}C_{1}}{2}\right)+\left[\delta(B_{2}C_{2})-\delta(B_{2})\delta(C_{2})\right]\ln\left(\frac{B_{1}C_{2}\delta(B_{2})+B_{2}C_{1}\delta(C_{2})}{2}\right)\nonumber\\
&+&\sum_{k=1}^{\infty}\frac{(-1)^{k}G_{k}}{\nu^{k}(1+z^{2})^{\frac{k}{2}}}\;,
\end{eqnarray}
where
\begin{equation}\label{46}
G_{k}=\left[1-\delta(B_{2}C_{2})\right]P_{k}
+\left[\delta(B_{2}C_{2})-\delta(B_{2})\delta(C_{2})\right]\frac{Q_{k}}{k}\;,
\end{equation}
with
\begin{equation}\label{47}
P_{k}=-\frac{1}{k}\left[\left(\frac{B_{1}}{B_{2}}\right)^{k}+\left(\frac{C_{1}}{C_{2}}\right)^{k}\right]\;,\quad
Q_{k}=\left(\frac{B_{1}C_{1}}{B_{1}C_{2}\delta(B_{2})+B_{2}C_{1}\delta(C_{2})}\right)^{k}\;.
\end{equation}

The uniform asymptotic expansions (\ref{43}) and (\ref{45}) can now be utilized to perform the analytic continuation of the
spectral zeta functions associated with the two regions. By adding and subtracting in (\ref{24}) $N$ leading terms of the asymptotic expansion (\ref{43})
the spectral zeta function in region $I$ can be written as
\begin{equation}\label{48}
\zeta_{I}(s,a)={\cal Z}_{I}(s,a)+\sum_{i=-1}^{N}{\cal A}^{I}_{i}(s,a)\;.
\end{equation}
The function ${\cal Z}_{I}(s,a)$ is analytic in the region $\Re(s)>(d-N-1)/2$ and has the integral representation
\begin{eqnarray}\label{49}
{\cal Z}_{I}(s,a)&=&\frac{\sin\pi s}{\pi}\sum_{\nu}d(\nu)\nu^{-2s}\int_{0}^{\infty}z^{-2s}\Bigg[\frac{\partial}{\partial z}\ln\Omega^{I}_{\nu}(i\nu z,a)
-\nu\sqrt{1+z^{2}}a-\left[1-\delta(A_{2}B_{2})\right]\ln\left(\frac{A_{2}B_{2}}{2}\right)\nonumber\\
&-&\left[1-\delta(A_{2}B_{2})-\delta(A_{2})\delta(B_{2})\right]\ln\left(\nu\sqrt{1+z^{2}}\right)
-\delta(A_{2})\delta(B_{2})\ln\left(\frac{A_{1}B_{1}}{2}\right)\nonumber\\
&-&\left[\delta(A_{2}B_{2})-\delta(A_{2})\delta(B_{2})\right]\ln\left(-\frac{A_{1}B_{2}\delta(A_{2})+A_{2}B_{1}\delta(B_{2})}{2}\right)
-\sum_{k=1}^{N}\frac{F_{k}}{\nu^{k}(1+z^{2})^{\frac{k}{2}}}\Bigg]\diff z\;.
\end{eqnarray}
The remaining terms in (\ref{49}), namely ${\cal A}^{I}_{i}(s,a)$, represent meromorphic functions of $s\in\mathbb{C}$ possesing only simple poles.
These terms can be expressed in terms of the spectral zeta function of the manifold $N$
\begin{equation}
\zeta_{N}(s)=\sum_{\nu}d(\nu)\nu^{-2s}\;,
\end{equation}
valid for $\Re(s)>d/2$, and have the explicit form
\begin{equation}\label{50}
{\cal A}^{I}_{-1}(s,a)=\frac{a}{2\sqrt{\pi}\Gamma(s)}\Gamma\left(s-\frac{1}{2}\right)\zeta_{N}\left(s-\frac{1}{2}\right)\;,
\end{equation}
\begin{equation}\label{51}
{\cal A}^{I}_{0}(s,a)=\frac{1}{2}\left[1-\delta(A_{2}B_{2})-\delta(A_{2})\delta(B_{2})\right]\zeta_{N}(s)\;,
\end{equation}
and, for $i\geq 1$,
\begin{equation}\label{52}
{\cal A}^{I}_{i}(s,a)=-\frac{F_i}{\Gamma\left(\frac{i}{2}\right)\Gamma(s)}\Gamma\left(s+\frac{i}{2}\right)\zeta_{N}\left(s+\frac{i}{2}\right)\;.
\end{equation}

The spectral zeta function in region $II$ can be written in a similar fashion
\begin{equation}\label{53}
\zeta_{II}(s,a)={\cal Z}_{II}(s,a)+\sum_{i=-1}^{N}{\cal A}^{II}_{i}(s,a)\;,
\end{equation}
where
\begin{eqnarray}\label{54}
{\cal Z}_{II}(s,a)&=&\frac{\sin\pi s}{\pi}\sum_{\nu}d(\nu)\nu^{-2s}\int_{0}^{\infty}z^{-2s}\Bigg[\frac{\partial}{\partial z}\ln\Omega^{II}_{\nu}(i\nu z,a)
-\nu\sqrt{1+z^{2}}(L-a)-\left[1-\delta(B_{2}C_{2})\right]\ln\left(\frac{B_{2}C_{2}}{2}\right)\nonumber\\
&-&\left[1-\delta(B_{2}C_{2})-\delta(B_{2})\delta(C_{2})\right]\ln\left(\nu\sqrt{1+z^{2}}\right)
-\delta(B_{2})\delta(C_{2})\ln\left(\frac{B_{1}C_{1}}{2}\right)\nonumber\\
&-&\left[\delta(B_{2}C_{2})-\delta(B_{2})\delta(C_{2})\right]\ln\left(\frac{B_{1}C_{2}\delta(B_{2})+B_{2}C_{1}\delta(C_{2})}{2}\right)
-\sum_{k=1}^{N}\frac{(-1)^{k}G_{k}}{\nu^{k}(1+z^{2})^{\frac{k}{2}}}\Bigg]\diff z\;,
\end{eqnarray}
is analytic for $\Re(s)>(d-N-1)/2$, and
\begin{equation}\label{55}
{\cal A}^{II}_{-1}(s,a)=\frac{L-a}{2\sqrt{\pi}\Gamma(s)}\Gamma\left(s-\frac{1}{2}\right)\zeta_{N}\left(s-\frac{1}{2}\right)\;,
\end{equation}
\begin{equation}\label{56}
{\cal A}^{II}_{0}(s,a)=\frac{1}{2}\left[1-\delta(B_{2}C_{2})-\delta(B_{2})\delta(C_{2})\right]\zeta_{N}(s)\;,
\end{equation}
\begin{equation}\label{57}
{\cal A}^{II}_{i}(s,a)=\frac{(-1)^{i+1}G_i}{\Gamma\left(\frac{i}{2}\right)\Gamma(s)}\Gamma\left(s+\frac{i}{2}\right)\zeta_{N}\left(s+\frac{i}{2}\right)\;,
\end{equation}
for $i\geq 1$, are meromorphic functions of $s\in\mathbb{C}$. The expressions (\ref{48}) and (\ref{53}) represent the analytic continuation
of the spectral zeta functions in region $I$ and $II$ and will be used to compute the Casimir energy and corresponding force on
the piston.

\section{Zero modes on the manifold $N$}\label{sec4}

The analytic continuation of $\zeta_{I}(s,a)$ and $\zeta_{II}(s,a)$ presented in the previous Section was performed under the tacit assumption that
the Laplacian $\Delta_{N}$ acting on functions defined on the manifold $N$ does not possess zero modes. Here, we drop that assumption and analyze the case in which
$\nu=0$ is an eiganvelue of $\Delta_{N}$ with multiplicity $d(0)$. Under these circumstances the process of analytic continuation
of the spectral zeta functions needs to be slightly modified since the large-$\nu$ asymptotic expansions used before have to be replaced with different ones.

The integral representation of the spectral zeta functions $\zeta_{i}(s,a)$ can be rewritten in a form that separates the contribution of the zero modes from the
rest, more precisely
\begin{equation}\label{58}
\zeta_{i}(s,a)=\frac{d(0)}{2\pi i}\int_{\gamma_{i}}\kappa^{-2s}\frac{\partial}{\partial \kappa}\ln\Omega^{i}_{0}(\kappa,a)\diff\kappa
+\frac{1}{2\pi i}\sum_{\nu}d(\nu)\int_{\gamma_{i}}\kappa^{-2s}\frac{\partial}{\partial \kappa}\ln\Omega^{i}_{\nu}(\kappa,a)\diff\kappa\;.
\end{equation}
Since the analytic continuation of the integral corresponding to the non-vanishing modes has been developed in the previous Section, it will not be repeated here.
We will be concerned, instead, with the analytic continuation of the first integral in (\ref{58}). For $\nu=0$, the differential equation describing the dynamics of the
scalar field is
\begin{equation}\label{59}
  \left(-\frac{\diff^{2}}{\diff x^{2}}-\alpha_{i}^{2}\right)h_{i}(x)=0\;,
\end{equation}
endowed with the boundary conditions (\ref{10}) in region $I$ and (\ref{11}) in region $II$. The boundary value problem consisting of
(\ref{59}) with the boundary conditions (\ref{10}) provides the following implicit equation for the eigenvalues $\alpha_{I}$ in region $I$
\begin{equation}\label{60}
  \Omega^{I}_{0}(\alpha,a)=\left(\frac{A_{1}B_{1}}{\alpha_{I}}-A_{2}B_{2}\alpha_{I}\right)\sin(\alpha_{I} a)-\left(A_{1}B_{2}+A_{2}B_{1}\right)\cos(\alpha_{I} a)=0\;.
\end{equation}
The solution of (\ref{59}) coupled with the conditions (\ref{11}) lead, in region $II$, to an implicit equation for the eigenvalues $\alpha_{II}$
\begin{equation}\label{61}
  \Omega^{II}_{0}(\alpha,a)=\left(\frac{B_{1}C_{1}}{\alpha_{II}}-B_{2}C_{2}\alpha_{II}\right)\sin\left[\alpha_{II}(L- a)\right]+\left(B_{1}C_{2}+B_{2}C_{1}\right)\cos\left[\alpha_{II}(L- a)\right]=0\;.
\end{equation}
It is not very difficult to verify that when the conditions (\ref{21}) and (\ref{21a}) are satisfied the equation (\ref{60}) has only real zeroes.
Likewise, when the conditions (\ref{30}) and (\ref{31}) hold, then the equation (\ref{61}) has only real solutions as well.

By deforming the contour of integration $\gamma_{i}$ to the imaginary axis we can write the contribution to the spectral zeta functions in region $I$
and $II$ as
\begin{equation}\label{62}
  \zeta_{i}^{0}(s,a)=d(0)\frac{\sin\pi s}{\pi}\int_{0}^{1}z^{-2s}\frac{\partial}{\partial z}\ln \Omega_{i}^{0}(iz,a)\diff z
  +d(0)\frac{\sin\pi s}{\pi}\int_{1}^{\infty}z^{-2s}\frac{\partial}{\partial z}\ln \Omega_{i}^{0}(iz,a)\diff z\;,
\end{equation}
where the first integral is convergent for $\Re(s)<1$ and the second for $\Re(s)>1/2$. To perform the analytic continuation of $\zeta_{i}^{0}(s,a)$
to a region to the left of the line $\Re(s)=1/2$ we need to find the asymptotic expansion of $\ln \Omega_{i}^{0}(iz,a)$ valid for large values of $z$.
In region $I$ we have
\begin{eqnarray}\label{63}
  \Omega_{I}^{0}(iz,a)=\frac{e^{z a}}{2z}\left(A_{1}B_{1}-(A_{1}B_{2}+A_{2}B_{1})z+A_{2}B_{2}z^{2}\right)\left(1+\epsilon^{0}_{I}(z,a)\right)\;,
\end{eqnarray}
with $\epsilon^{0}_{I}(z,a)$ exponentially small terms. For the purpose of analytic continuation we need, however, the natural logarithm of (\ref{63})
which has the expression
\begin{eqnarray}\label{64}
 \ln\Omega_{I}^{0}(iz,a)&=&z a-\ln2+\left[1-\delta(A_{2}B_{2})-\delta(A_{2})\delta(B_{2})\right]\ln z
+\left[1-\delta(A_{2}B_{2})\right]\ln\left(A_{2}B_{2}\right)\nonumber\\
&+&\delta(A_{2})\delta(B_{2})\ln\left(A_{1}B_{1}\right)+\left[\delta(A_{2}B_{2})-\delta(A_{2})\delta(B_{2})\right]\ln\left[-A_{1}B_{2}\delta(A_{2})-A_{2}B_{1}\delta(B_{2})\right]\nonumber\\
&+&\left[1-\delta(A_{2}B_{2})\right]\ln\left[1-\frac{A_{1}B_{2}+A_{2}B_{1}}{A_{2}B_{2}}\frac{1}{z}+\frac{A_{1}B_{1}}{A_{2}B_{2}}\frac{1}{z^{2}}\right]\nonumber\\
&+&\left[\delta(A_{2}B_{2})-\delta(A_{2})\delta(B_{2})\right]\ln\left[1-\frac{A_{1}B_{1}}{A_{1}B_{2}\delta(A_{2})+A_{2}B_{1}\delta(B_{2})}\frac{1}{z}\right]
+\ln\left[1+\epsilon^{0}_{I}(z,a)\right]\;.
\end{eqnarray}
By exploiting asymptotic expansions similar to the ones in (\ref{36}) and (\ref{40}) we obtain the large-$z$ asymptotic expansion of $\ln\Omega_{i}^{0}(iz,a)$
in the form
\begin{eqnarray}\label{65}
\ln\Omega_{I}^{0}(iz,a)&\sim&za-\ln2+\left[1-\delta(A_{2}B_{2})-\delta(A_{2})\delta(B_{2})\right]\ln z
+\left[1-\delta(A_{2}B_{2})\right]\ln\left(A_{2}B_{2}\right)\nonumber\\
&+&\delta(A_{2})\delta(B_{2})\ln\left(A_{1}B_{1}\right)+\left[\delta(A_{2}B_{2})-\delta(A_{2})\delta(B_{2})\right]\ln\left[-A_{1}B_{2}\delta(A_{2})-A_{2}B_{1}\delta(B_{2})\right]\nonumber\\
&+&\sum_{k=1}^{\infty}\frac{F_{k}}{z^{k}}\;,
\end{eqnarray}
where the terms $E_{k}$ are given in (\ref{43a}) exponentially small contributions have been omitted.

In region $II$ we follow an analogous procedure to get the following large-$z$ asymptotic expansion of $\ln\Omega_{II}^{0}(iz,a)$
\begin{eqnarray}\label{66}
\ln\Omega_{II}^{0}(iz,a)&\sim&z(L-a)-\ln2+\left[1-\delta(B_{2}C_{2})-\delta(B_{2})\delta(C_{2})\right]\ln z
+\left[1-\delta(B_{2}C_{2})\right]\ln\left(B_{2}C_{2}\right)\nonumber\\
&+&\delta(B_{2})\delta(C_{2})\ln\left(B_{1}C_{1}\right)+\left[\delta(B_{2}C_{2})-\delta(B_{2})\delta(C_{2})\right]\ln\left[B_{1}C_{2}\delta(B_{2})+B_{2}C_{1}\delta(C_{2})\right]\nonumber\\
&+&\sum_{k=1}^{\infty}\frac{(-1)^{k}G_{k}}{z^{k}}\;,
\end{eqnarray}
where, once again, exponentially small terms have been discarded and the coefficients $G_{k}$ are defined in (\ref{46}).

We proceed as before by adding and subtracting from the second integral in (\ref{62}) $N$ leading terms of the asymptotic expansions (\ref{65}) in region $I$ and (\ref{66}) in region $I$.
Hence, the analytically continued expression of $\zeta_{I}^{0}(s,a)$ is found to be
\begin{equation}\label{67}
  \zeta_{I}^{0}(s,a)={\cal Z}_{I}^{0}(s,a)+d(0)\frac{\sin\pi s}{\pi}\left[\frac{a}{2s-1}+\frac{1-\delta(A_{2}B_{2})-\delta(A_{2})\delta(B_{2})}{2s}
  -\sum_{k=1}^{N}\frac{k F_{k}}{2s+k}\right]\;,
\end{equation}
and the one for $\zeta_{II}^{0}(s,a)$ is, instead,
\begin{equation}\label{68}
  \zeta_{II}^{0}(s,a)={\cal Z}_{II}^{0}(s,a)+d(0)\frac{\sin\pi s}{\pi}\left[\frac{L-a}{2s-1}+\frac{1-\delta(B_{2}A_{2})-\delta(B_{2})\delta(A_{2})}{2s}
  -\sum_{k=1}^{N}\frac{ (-1)^{k}k G_{k}}{2s+k}\right]\;,
\end{equation}
where the results in square parenthesis are obtained from the elementary integration of the asymptotic terms. The functions ${\cal Z}_{I}^{0}(s,a)$ and ${\cal Z}_{II}^{0}(s,a)$
are analytic in $s$ in the region $\Re(s)>-(N+1)/2$ and have the integral representation
\begin{eqnarray}\label{69}
  {\cal Z}_{I}^{0}(s,a)&=&d(0)\frac{\sin\pi s}{\pi}\int_{0}^{\infty}z^{-2s}\frac{\partial}{\partial z}\Bigg[\ln\Omega_{I}^{0}(iz,a)-\Theta(z-1)\Bigg(za-\ln2\nonumber\\
&+&\left[1-\delta(A_{2}B_{2})-\delta(A_{2})\delta(B_{2})\right]\ln z+\left[1-\delta(A_{2}B_{2})\right]\ln\left(A_{2}B_{2}\right)
+\delta(A_{2})\delta(B_{2})\ln\left(A_{1}B_{1}\right)\nonumber\\
&+&\left[\delta(A_{2}B_{2})-\delta(A_{2})\delta(B_{2})\right]\ln\left[-A_{1}B_{2}\delta(A_{2})-A_{2}B_{1}\delta(B_{2})\right]
+\sum_{k=1}^{N}\frac{F_{k}}{z^{k}}\Bigg)\Bigg]\diff z\;,
\end{eqnarray}
and
\begin{eqnarray}\label{70}
  {\cal Z}_{II}^{0}(s,a)&=&d(0)\frac{\sin\pi s}{\pi}\int_{0}^{\infty}z^{-2s}\frac{\partial}{\partial z}\Bigg[\ln\Omega_{II}^{0}(iz,a)-\Theta(z-1)\Bigg(z(L-a)-\ln2\nonumber\\
&+&\left[1-\delta(B_{2}C_{2})-\delta(B_{2})\delta(C_{2})\right]\ln z+\left[1-\delta(B_{2}C_{2})\right]\ln\left(B_{2}C_{2}\right)
+\delta(B_{2})\delta(C_{2})\ln\left(B_{1}C_{1}\right)\nonumber\\
&+&\left[\delta(B_{2}C_{2})-\delta(B_{2})\delta(C_{2})\right]\ln\left[B_{1}C_{2}\delta(B_{2})+B_{2}C_{1}\delta(C_{2})\right]
+\sum_{k=1}^{N}\frac{(-1)^{k}G_{k}}{z^{k}}\Bigg)\Bigg]\diff z\;,
\end{eqnarray}
with $\Theta(x)$ denoting the Heaviside step-function. The expressions obtained in (\ref{67}) and (\ref{68}) represent the contributions
to the spectral zeta function that are present when the Laplacian $\Delta_{N}$ on the manifold $N$ possesses zero modes.

\section{casimir energy and force}\label{sec5}

According to the formula displayed in (\ref{7}), the Casimir energy associated with the piston configuration is computed by
performing the limit as $s\to-1/2$ of the spectral zeta function $\zeta(s)$. By setting $N=D$ in the results (\ref{48}) and (\ref{53}) we obtain a representation
of $\zeta(s,a)$ valid in the region $-1<\Re(s)<1$
\begin{eqnarray}\label{71}
\zeta(s,a)=\zeta_{I}(s,a)+\zeta_{II}(s,a)&=&{\cal Z}_{I}(s,a)+{\cal Z}_{II}(s,a)+\frac{L}{2\sqrt{\pi}\Gamma(s)}\Gamma\left(s-\frac{1}{2}\right)\zeta_{N}\left(s-\frac{1}{2}\right)\nonumber\\
&+&\frac{1}{2}\left[2-\delta(A_{2}B_{2})-\delta(B_{2}C_{2})-\delta(A_{2})\delta(B_{2})-\delta(B_{2})\delta(C_{2})\right]\zeta_{N}(s)\nonumber\\
&-&\sum_{k=1}^{D}\frac{F_{k}+(-1)^{k}G_{k}}{\Gamma\left(\frac{k}{2}\right)\Gamma(s)}\Gamma\left(s+\frac{k}{2}\right)\zeta_{N}\left(s+\frac{k}{2}\right)\;.
\end{eqnarray}
One can extract the Casimir energy from the above expression first by substituting $s=\varepsilon-1/2$ and then by taking the limit $\varepsilon\to 0$.
This limit, however, also shows the meromorphic structure of $\zeta(s)$ in the neighborhood of $s=-1/2$. The pole structure of $\zeta(s)$
is intimately related to the one of $\zeta_{N}(s)$ which, according to the general theory of spectral zeta functions, is \cite{gilkey95,kirsten01}
\begin{eqnarray}\label{72}
\zeta_{N}(\varepsilon -n)&=&\zeta_{N}(-n)+\varepsilon\zeta^{\prime}_{N}(-n)+O(\varepsilon^{2})\;,\\
\zeta_{N}\left(\varepsilon+\frac{d-k}{2}\right)&=&\frac{1}{\varepsilon}\textrm{Res}\,\zeta_{N}\left(\frac{d-k}{2}\right)+\textrm{FP}\,\zeta_{N}\left(\frac{d-k}{2}\right)+O(\varepsilon)\;,\\
\zeta_{N}\left(\varepsilon-\frac{2n+1}{2}\right)&=&\frac{1}{\varepsilon}\textrm{Res}\,\zeta_{N}\left(-\frac{2n+1}{2}\right)+\textrm{FP}\,\zeta_{N}\left(\frac{2n+1}{2}\right)+O(\varepsilon)\;,
\end{eqnarray}
where $n\in\mathbb{N}_{0}$ and $k=\{0,\ldots,d-1\}$. We would like to mention that the residues of the spectral zeta function $\zeta_{N}(s)$ are related to
the geometry of the manifold $N$ as they are proportional to the coefficients of the heat kernel asymptotic expansion \cite{gilkey95,kirsten01}. More precisely, one has
\begin{equation}\label{72a}
  \Gamma\left(\frac{d-k}{2}\right)\textrm{Res}\,\zeta_{N}\left(\frac{d-k}{2}\right)=A^{N}_{\frac{k}{2}}\;,\quad
  \Gamma\left(-\frac{2n+1}{2}\right)\textrm{Res}\,\zeta_{N}\left(-\frac{2n+1}{2}\right)=A^{N}_{\frac{d+2n+1}{2}}\;.
\end{equation}

Since ${\cal Z}_{I}(s,a)$ and ${\cal Z}_{II}(s,a)$ are analytic functions for $-1<\Re(s)<1$ the value $-1/2$ can be simply substituted for $s$. For the other terms in (\ref{71}) we have
\begin{equation}\label{73}
  \frac{L}{2\sqrt{\pi}\Gamma\left(\varepsilon-\frac{1}{2}\right)}\Gamma(\varepsilon-1)\zeta_{N}(\varepsilon-1)=\frac{L\,\zeta_{N}(-1)}{4\pi\varepsilon}
  +\frac{L}{4\pi}\left[\zeta'_{N}(-1)+(2\ln 2-1)\zeta_{N}(-1)\right]+O(\varepsilon)\;,
\end{equation}
and
\begin{eqnarray}\label{74}
  \lefteqn{\frac{1}{2}\left[2-\delta(A_{2}B_{2})-\delta(B_{2}C_{2})-\delta(A_{2})\delta(B_{2})-\delta(B_{2})\delta(C_{2})\right]\zeta_{N}\left(\varepsilon-\frac{1}{2}\right)}\nonumber\\
  &&=\frac{1}{2\varepsilon}\left[2-\delta(A_{2}B_{2})-\delta(B_{2}C_{2})-\delta(A_{2})\delta(B_{2})-\delta(B_{2})\delta(C_{2})\right]\textrm{Res}\,\zeta_{N}\left(-\frac{1}{2}\right)\nonumber\\
  &&+\frac{1}{2}\left[2-\delta(A_{2}B_{2})-\delta(B_{2}C_{2})-\delta(A_{2})\delta(B_{2})-\delta(B_{2})\delta(C_{2})\right]\textrm{FP}\,\zeta_{N}\left(-\frac{1}{2}\right)+O(\varepsilon)\;.
\end{eqnarray}
For the last term in (\ref{71}) we have, instead, for $k=1$ the expansion
\begin{equation}\label{75}
  -\frac{F_{1}-G_{1}}{\sqrt{\pi}\Gamma\left(\varepsilon-\frac{1}{2}\right)}\Gamma\left(\varepsilon\right)\zeta_{N}\left(\varepsilon\right)
  =\frac{F_{1}-G_{1}}{2\pi\varepsilon}\zeta_{N}(0)+\frac{F_{1}-G_{1}}{2\pi}\left[\zeta'_{N}(0)+2(\ln 2-1)\zeta_{N}(0)\right]+O(\varepsilon)\;,
\end{equation}
and, for $k=\{2,\ldots,D\}$,
\begin{eqnarray}\label{76}
\lefteqn{-\frac{F_{k}+(-1)^{k}G_{k}}{\Gamma\left(\frac{k}{2}\right)\Gamma\left(\varepsilon-\frac{1}{2}\right)}\Gamma\left(\varepsilon+\frac{k-1}{2}\right)\zeta_{N}\left(\varepsilon+\frac{k-1}{2}\right)
=\frac{F_{k}+(-1)^{k}G_{k}}{2\sqrt{\pi}\Gamma\left(\frac{k}{2}\right)\varepsilon}\Gamma\left(\frac{k-1}{2}\right)\textrm{Res}\,\zeta_{N}\left(\frac{k-1}{2}\right)}\nonumber\\
&&+\frac{F_{k}+(-1)^{k}G_{k}}{2\sqrt{\pi}\Gamma\left(\frac{k}{2}\right)}\Gamma\left(\frac{k-1}{2}\right)\left[\textrm{FP}\,\zeta_{N}\left(\frac{k-1}{2}\right)+
\left(2-\gamma-2\ln 2+\Psi\left(\frac{k-1}{2}\right)\right)\textrm{Res}\,\zeta_{N}\left(\frac{k-1}{2}\right)\right]+O(\varepsilon)\;.\;\;
\end{eqnarray}

The results (\ref{73})-(\ref{76}) obtained above provide the Casimir energy of the piston thanks to the formula (\ref{8}). In more details one has
\begin{eqnarray}\label{77}
\lefteqn{E_{\textrm{Cas}}(a)=\frac{1}{2}\left(\frac{1}{\varepsilon}+\ln\mu^{2}\right)\Bigg[\frac{1}{2}\left[2-\delta(A_{2}B_{2})-\delta(B_{2}C_{2})-\delta(A_{2})\delta(B_{2})-\delta(B_{2})\delta(C_{2})\right]\textrm{Res}\,\zeta_{N}\left(-\frac{1}{2}\right)}\nonumber\\
&+&\frac{L}{4\pi}\zeta_{N}(-1)+\frac{F_{1}-G_{1}}{2\pi}\zeta_{N}(0)+\sum_{k=2}^{D}\frac{F_{k}+(-1)^{k}G_{k}}{2\sqrt{\pi}\Gamma\left(\frac{k}{2}\right)}\Gamma\left(\frac{k-1}{2}\right)\textrm{Res}\,\zeta_{N}\left(\frac{k-1}{2}\right)\Bigg]\nonumber\\
&+&\frac{1}{2}{\cal Z}_{I}\left(-\frac{1}{2},a\right)+\frac{1}{2}{\cal Z}_{II}\left(-\frac{1}{2},a\right)+\frac{L}{8\pi}\left[\zeta'_{N}(-1)+(2\ln 2-1)\zeta_{N}(-1)\right]\nonumber\\
&+&\frac{1}{4}\left[2-\delta(A_{2}B_{2})-\delta(B_{2}C_{2})-\delta(A_{2})\delta(B_{2})-\delta(B_{2})\delta(C_{2})\right]\textrm{FP}\,\zeta_{N}\left(-\frac{1}{2}\right)\nonumber\\
&+&\frac{F_{1}-G_{1}}{4\pi}\left[\zeta'_{N}(0)+2(\ln 2-1)\zeta_{N}(0)\right]\nonumber\\
&+&\sum_{k=2}^{D}\frac{F_{k}+(-1)^{k}G_{k}}{4\sqrt{\pi}\Gamma\left(\frac{k}{2}\right)}\Gamma\left(\frac{k-1}{2}\right)\left[\textrm{FP}\,\zeta_{N}\left(\frac{k-1}{2}\right)+
\left(2-\gamma-2\ln 2+\Psi\left(\frac{k-1}{2}\right)\right)\textrm{Res}\,\zeta_{N}\left(\frac{k-1}{2}\right)\right]+O(\varepsilon)\;.\nonumber\\
\end{eqnarray}
As it is to be expected the Casimir energy associated with the piston is, in general, not a well defined quantity \cite{bordag09}.
The explicit result (\ref{77}) shows that the ambiguity in the energy is essentially dependent on the geometry of the manifold $N$.
In fact, the terms responsible for the ambiguity, namely the ones multiplying $(1/\varepsilon+\ln\mu^{2})$, are proportional to the
the heat kernel coefficients $A^{N}_{(D-k)/2}$ with $k=\{-2,\ldots,D\}$. This type of ambiguity in the Casimir energy is always
found in higher dimensional piston configurations \cite{bea13,fucci12}.

Despite the intrinsic ambiguity that is present in the Casimir energy, the Casimir force acting on the piston is a well defined quantity.
By applying the definition (\ref{9}) to the result for the Casimir energy in (\ref{77}) we obtain the following expression
for the Casimir force
\begin{eqnarray}\label{78}
F_{\textrm{Cas}}(a)=-\frac{1}{2}{\cal Z}^{\prime}_{I}\left(-\frac{1}{2},a\right)+\frac{1}{2}{\cal Z}^{\prime}_{II}\left(-\frac{1}{2},a\right)\;,
\end{eqnarray}
with the prime denoting differentiation with respect to the variable $a$. The results for the Casimir energy (\ref{77}) and the Casimir force (\ref{78})
are very general as they are valid for any smooth compact Riemannian manifold $N$ and for all values of the coefficients $(A_{1},A_{2},B_{1},B_{2},C_{1},C_{2})$
satisfying the conditions (\ref{21}), (\ref{21a}), (\ref{30}), and (\ref{31}). In order to obtain more explicit results one has to specify
the manifold $N$ and the values of the coefficients describing the boundary conditions.

We would like to conclude this Section by considering the contribution to the Casimir energy and force due to the presence of zero
modes associated with the Laplacian on the manifold $N$. By setting $N=D$ and $s=\varepsilon-1/2$ in the sum of (\ref{67}) and (\ref{68}) we obtain, as $\varepsilon\to0$,
\begin{eqnarray}\label{79}
E^{0}_{\textrm{Cas}}(a)&=&\frac{d(0)}{4\pi}\left(\frac{1}{\varepsilon}+\ln\mu^{2}\right)\left(F_{1}-G_{1}\right)+\frac{1}{2}{\cal Z}^{0}_{I}\left(-\frac{1}{2},a\right)+\frac{1}{2}{\cal Z}^{0}_{II}\left(-\frac{1}{2},a\right)\nonumber\\
&+&\frac{d(0)}{2\pi}\left[\frac{L}{2}+2-\delta(A_{2}B_{2})-\delta(B_{2}C_{2})-\delta(A_{2})\delta(B_{2})-\delta(B_{2})\delta(C_{2})\right]\nonumber\\
&+&\frac{d(0)}{2\pi}\sum_{k=2}^{D}\frac{k[F_{k}+(-1)^{k}G_{k}]}{k-1}+O(\varepsilon)\;.
\end{eqnarray}
It is clear from the above result that the contribution to the Casimir energy of the piston coming from the zero modes is not well defined.
This is a feature that is also been observed in (\ref{77}). However, unlike the ambiguity present in (\ref{77}) which is dependent on
to the geometry of the manifold $N$, the one in (\ref{79}) is subject to the particular boundary conditions imposed. The contribution of the zero modes
to the Casimir force on the piston, is obtained by differentiating (\ref{79}) with respect to $a$. In particular we have
\begin{equation}\label{79a}
F_{\textrm{Cas}}^{0}(a)=-\frac{1}{2}\left({\cal Z}^{0}_{I}\right)^{\prime}\left(-\frac{1}{2},a\right)+\frac{1}{2}\left({\cal Z}^{0}_{II}\right)^{\prime}\left(-\frac{1}{2},a\right)\;.
\end{equation}

\section{Specific Piston and Boundary Conditions}\label{sec6}

As we have mentioned in the previous Section explicit results for the Casimir force, $F_{\textrm{Cas}}(a)$, as a function of the position
of the piston $a$ can be obtained from (\ref{78}) once
the manifold $N$ has been selected and specific values have been assigned to the parameters $(A_{1},A_{2},B_{1},B_{2},C_{1},C_{2})$
according to the conditions (\ref{21}), (\ref{21a}), (\ref{30}), and (\ref{31}). In this work, however, we are mainly interested in
studying the behavior of $F_{\textrm{Cas}}(a)$ {\it as the boundary conditions
change}, rather than focusing on $F_{\textrm{Cas}}(a)$ with fixed boundary conditions. Within this framework, the Casimir force on the piston
is regarded not only as a function of the position $a$ but also as a function of the six parameters describing the boundary conditions.
It is clear that a complete analysis of the behavior of $F_{\textrm{Cas}}(a)$ as the six parameters vary independently becomes a rather prohibitive task.
For this reason, we will consider simplified cases
in which some of the parameters are kept fixed and the remaining ones are either allowed to vary in suitable intervals or are dependent on
each other so that the actual number of independent parameters is reduced.

In this Section we focus on a piston configuration of length $L=1$ for which the piston itself $N$ is a $d$-dimensional sphere. In this case
the eigenvalues of the Laplacian on $N$ are explicitly
known and have the form
\begin{equation}\label{79}
\nu=\left(l+\frac{d-1}{2}\right)\;,
\end{equation}
where $l\in\mathbb{N}_{0}$. The eigenfunctions on $N$ are hyperspherical harmonics having degeneracy
\begin{equation}\label{80}
d(l)=(2l+d-1)\frac{(l+d-2)!}{l!(d-1)!}\;.
\end{equation}
By using the formulas (\ref{79}) and (\ref{80}) the spectral zeta function $\zeta_{N}(s)$ can be written as
\begin{equation}\label{81}
\zeta_{N}(s)=\sum_{l=0}^{\infty}(2l+d-1)\frac{(l+d-2)!}{l!(d-1)!}\left(l+\frac{d-1}{2}\right)^{-2s}\;,
\end{equation}
which, in turn, can be expressed as linear combination of Hurwitz zeta functions \cite{bordag96,bordag96a,fucci11,fucci11b}
\begin{equation}\label{82}
\zeta_{N}(s)=2\sum_{\alpha=0}^{d-1}e_{\alpha}\zeta_{H}\left(2s-\alpha-1,\frac{d-1}{2}\right)\;,
\end{equation}
where the coefficients $e_{\alpha}$ can be found according to the relation
\begin{equation}\label{83}
\frac{(l+d-2)!}{l!(d-1)!}=\sum_{\alpha=0}^{d-1}e_{\alpha}\left(l+\frac{d-1}{2}\right)^{\alpha}\;.
\end{equation}
For this particular piston configuration we analyze the Casimir force $F_{\textrm{Cas}}(a)$, found in (\ref{78}), by imposing Dirichlet or Neumann boundary conditions on one of
the edges of the piston configuration or on the piston itself and by keeping general boundary conditions on the remaining two.
It is important to point out here that the choice of Dirichlet or Neumann boundary conditions has only been made for definiteness. In fact, any other choice
that fixes two of the six independent parameters is obviously acceptable.
In all the examples below we assume that the piston $N$ has dimension $d=2$.

\paragraph{Dirichlet and Neumann boundary conditions at $x=0$.}
First, we consider the cases in which Dirichlet or Neumann boundary conditions are imposed to the left-end of the piston, namely at $x=0$,
and the remaining boundary conditions are described by two independent parameters $\alpha$ and $\beta$. For the case of Dirichlet boundary conditions at $x=0$,
the coefficients in (\ref{10}) and (\ref{11}) can be expressed as
\begin{equation}\label{84}
A_{1}=1\;,A_{2}=0\;, B_{1}=\sin\alpha\;, B_{2}=\cos\alpha\;, C_{1}=\sin\beta\;, C_{2}=\cos\beta\;.
\end{equation}
Here and in the rest of this paper, we assume, without loss of generality, that $\alpha\in[0,\pi)$ and $\beta\in[0,\pi)$.
According to the conditions (\ref{21}), in region $I$ we need to impose the inequalities
\begin{equation}\label{85}
\frac{\cot\alpha}{a}\geq 1\;,\quad\textrm{or}\quad \cot\alpha\leq 0\;,
\end{equation}
In region $II$ the conditions (\ref{30}) lead, instead, to the inequalities
\begin{equation}\label{86}
\{\cot\alpha \cot\beta\leq 0\;, \cot\alpha+\cot\beta\leq -1+a\}\;,\quad\textrm{or}\quad \{\cot\alpha\geq 0\;, \cot\beta\geq 0\}\;.
\end{equation}
The inequalities (\ref{85}) and (\ref{86}) are simultaneously satisfied for all $a\in[0,1]$ if the parameters $\alpha$ and $\beta$ belong to either of the regions
$\alpha\in[0,\pi/4]$ and $\beta\in[\pi-\arctan(1+\cot\alpha),\pi)$, or $\alpha\in[3\pi/4,\pi)$ and $\beta\in[-\arctan(1+\cot\alpha),\pi/2]$, or
$\alpha\in[0,\pi/4]$ and $\beta\in[0,\pi/2]$.

For the case of Neumann boundary conditions at $x=0$ the coefficients in (\ref{10}) and (\ref{11}) can, instead, be written as
\begin{equation}\label{87}
A_{1}=0\;,A_{2}=1\;, B_{1}=\sin\alpha\;, B_{2}=\cos\alpha\;, C_{1}=\sin\beta\;, C_{2}=\cos\beta\;.
\end{equation}
The conditions in (\ref{21a}) for region $I$ imply that the parameter $\alpha$ need to satisfy the inequality
\begin{equation}\label{88}
\cot\alpha<0\;,
\end{equation}
while the conditions (\ref{30}) in region $II$ lead to the same inequalities found in (\ref{86}).
In order for (\ref{88}) and (\ref{86}) to be satisfied for all $a\in[0,1]$ the parameters $\alpha$ and $\beta$ need to take their values in the region
$\alpha\in[3\pi/4,\pi)$ and $\beta\in[-\arctan(1+\cot\alpha),\pi/2]$.

\begin{figure}[]
\centering
\includegraphics[scale=0.60,trim=0cm 0cm 0cm 0cm, clip=true, angle=0]{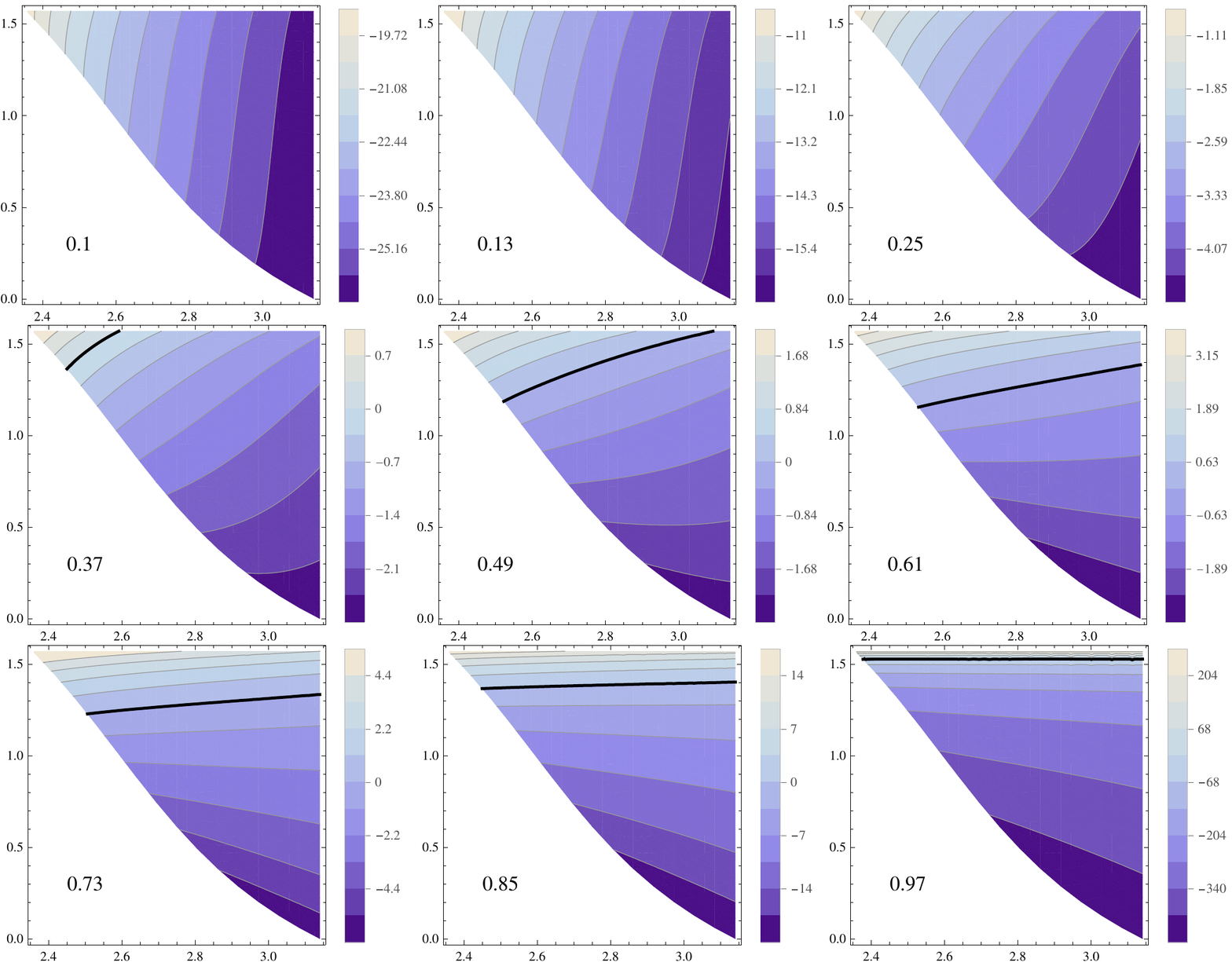}
\caption{Dirichlet boundary conditions at $x=0$. Each contour plot is obtained by fixing the value of the postion $a$, displayed in the bottom left corner, in the interval $[0,1]$.
The parameter $\alpha$ varies along the $x$-axis, while the parameter $\beta$ varies along the $y$-axis. The legends on the right provide the magnitude (in units for which $h=c=1$) and sign of the Casimir force on the piston.}\label{fig1}
\end{figure}

\begin{figure}[h!]
\centering
\includegraphics[scale=0.60]{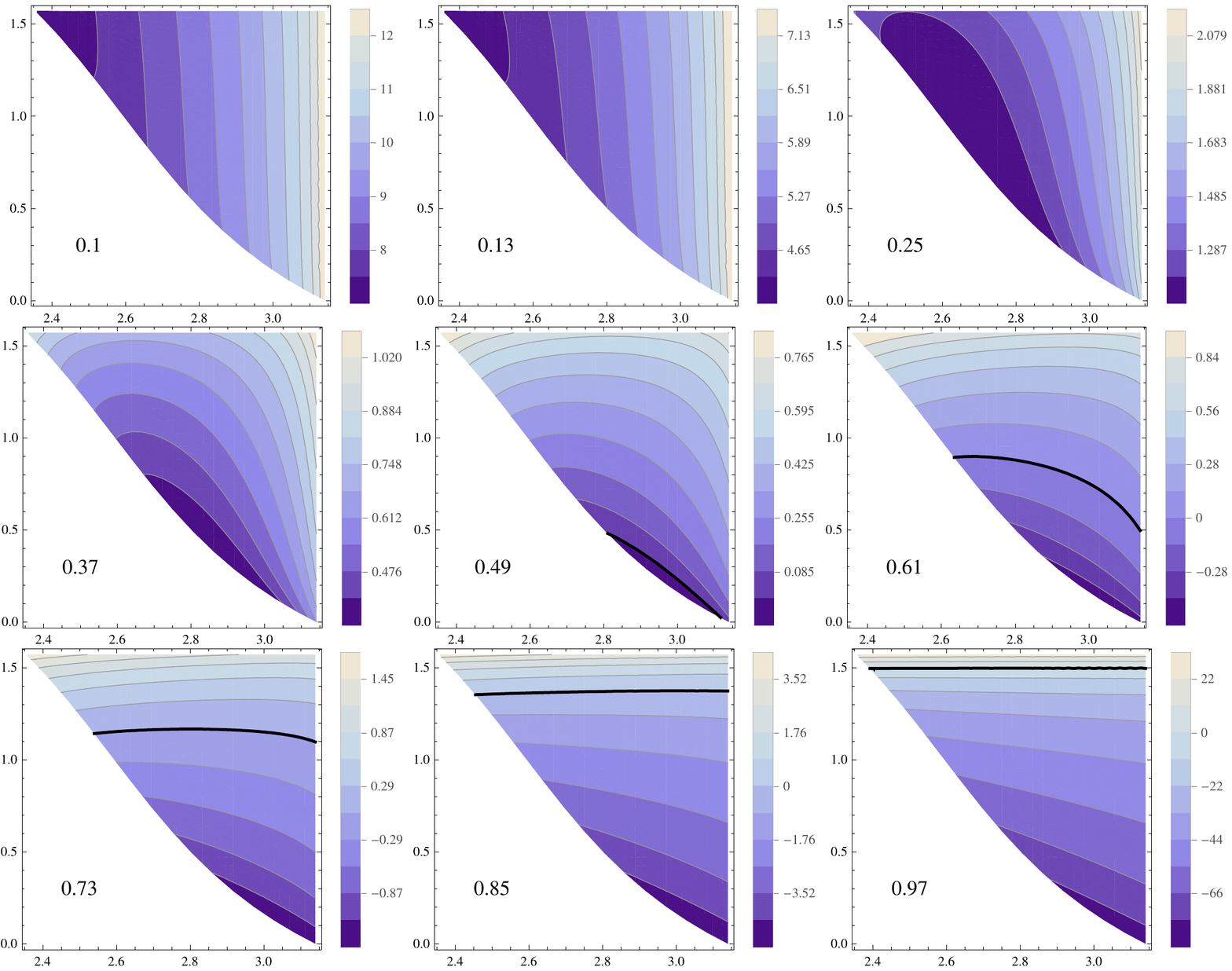}
\caption{Neumann boundary conditions at $x=0$. Each contour plot is obtained by fixing the value of the postion $a$, displayed in the bottom left corner, in the interval $[0,1]$.
The parameter $\alpha$ varies along the $x$-axis, while the parameter $\beta$ varies along the $y$-axis. The legends on the right provide the magnitude (in units for which $h=c=1$) and sign of the Casimir force on the piston.}\label{fig2}
\end{figure}

Contour plots of the Casimir
force on the piston (\ref{78}) for different values of the position of the piston $a$ are displayed in Figure \ref{fig1},
for the Dirichlet case, and in Figure \ref{fig2} for the Neumann case. For both the Dirichlet and Neumann cases the parameters
$\alpha$ and $\beta$ vary in the allowed region $\alpha\in[3\pi/4,\pi)$ and $\beta\in[-\arctan(1+\cot\alpha),\pi/2]$.
The curves displayed in bold represent the values of the pair $(\alpha,\beta)$
for which the Casimir force on the piston vanishes. The Casimir force is positive for values of $(\alpha,\beta)$ in the region above the
curves in bold, and it is negative for values of $(\alpha,\beta)$ in the region below them.

\paragraph{Dirichlet and Neumann boundary conditions on the piston at $x=a$}

As further examples, we consider the cases in which either Dirichlet or Neumann boundary conditions are imposed on the piston itself.
The remaining boundary conditions, at the two ends of the piston configuration, are assumed, as in the previous examples,
to be described by the two independent parameters $\alpha$ and $\beta$. When Dirichlet boundary conditions are imposed on the piston at $x=a$,
one can write the coefficients in (\ref{10}) and (\ref{11}) as
\begin{equation}\label{89}
A_{1}=\sin\alpha\;,A_{2}=\cos\alpha\;, B_{1}=1\;, B_{2}=0\;, C_{1}=\sin\beta\;, C_{2}=\cos\beta\;.
\end{equation}
By imposing the conditions (\ref{21}) in region $I$ we obtain the inequalities in (\ref{85}) while the conditions (\ref{30}) in region $II$ lead to
\begin{equation}\label{90}
\frac{\cot\beta}{1-a}\leq -1\;,\quad\textrm{or}\quad \cot\beta\geq 0\;.
\end{equation}
The inequalities (\ref{85}) and (\ref{90}) are satisfied for all $a\in[0,1]$ if $\alpha\in[0,\pi/4]\cup[\pi/2,\pi)$ and $\beta\in[3\pi/2,\pi)\cup(0,\pi/2]$.

When Neumann boundary conditions are imposed on the piston at $x=a$, the coefficients in (\ref{10}) and (\ref{11}) can be represented as
\begin{equation}\label{91}
A_{1}=\sin\alpha\;,A_{2}=\cos\alpha\;, B_{1}=0\;, B_{2}=1\;, C_{1}=\sin\beta\;, C_{2}=\cos\beta\;.
\end{equation}
The constraints (\ref{21a}) and (\ref{31}) imply that the following inequalities
\begin{equation}\label{92}
\frac{\cot\alpha}{a}<0\;,\quad \textrm{and}\quad \frac{\cot\beta}{1-a}>0\;.
\end{equation}
need to be satisfied for all $a\in[0,1]$ in region $I$ and region $II$, respectively. This is the case if the parameters $\alpha$ and $\beta$
are allowed to vary in the region $\alpha\in(\pi/2,\pi)$ and $\beta\in(0,\pi/2)$.

\begin{figure}[h!]
\centering
\includegraphics[scale=0.60,trim=0cm 0cm 0cm 0cm, clip=true, angle=0]{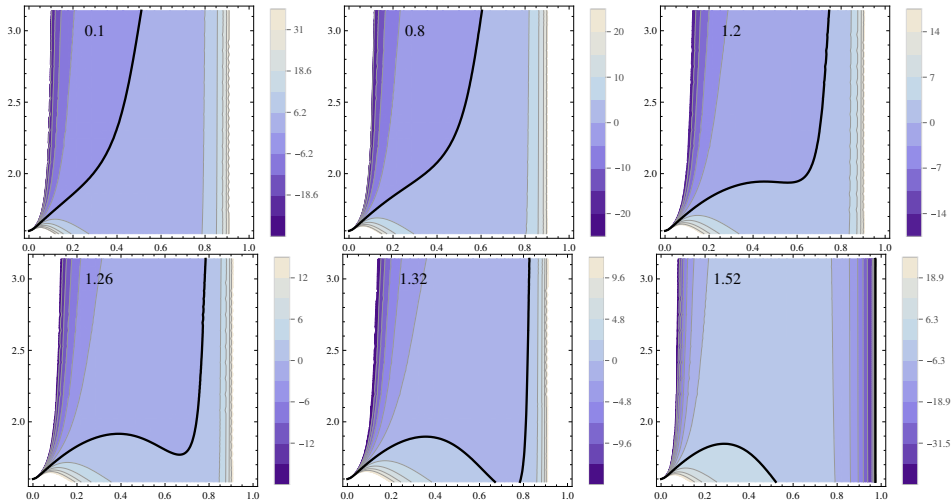}
\caption{Dirichlet boundary conditions at $x=a$. Each contour plot is obtained by fixing the value of the parameter $\beta$, displayed in the upper left corner, in the interval $(0,\pi/2)$.
The parameter $a$ varies along the $x$-axis, while the parameter $\alpha$ varies along the $y$-axis. The legends on the right provide the magnitude (in units for which $h=c=1$) and sign of the Casimir force on the piston.}\label{fig3}
\end{figure}

\begin{figure}[h!]
\centering
\includegraphics[scale=0.60,trim=0cm 0cm 0cm 0cm, clip=true, angle=0]{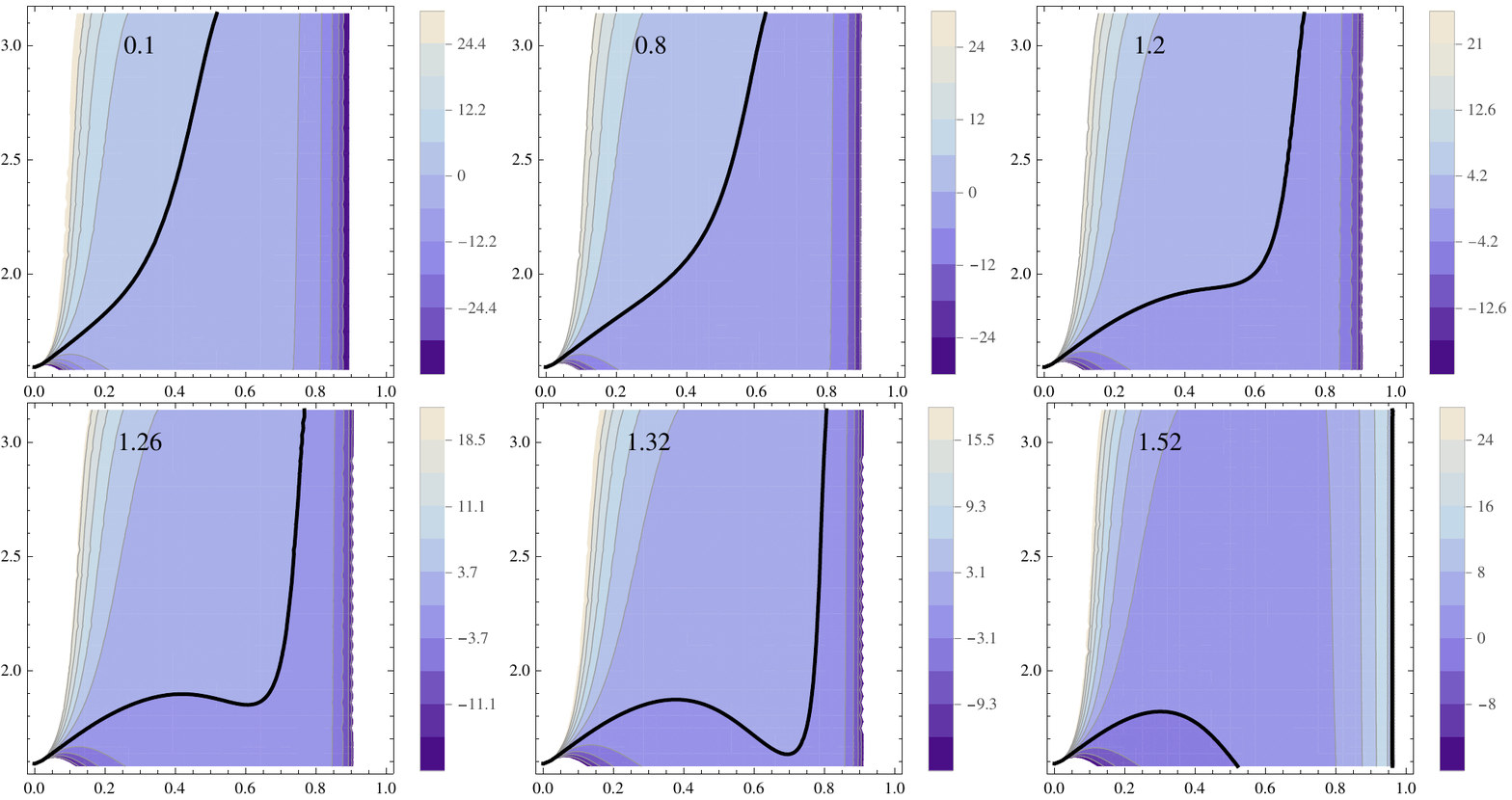}
\caption{Neumann boundary conditions at $x=a$. Each contour plot is obtained by fixing the value of the parameter $\beta$, displayed in the upper left corner, in the interval $(0,\pi/2)$. The parameter $a$ varies along the $x$-axis, while the parameter $\alpha$ varies along the $y$-axis. The legends on the right provide the magnitude (in units for which $h=c=1$) and sign of the Casimir force on the piston.}\label{fig4}
\end{figure}

Contour plots of the Casimir
force on the piston (\ref{78}) for different values of the parameter $\beta$ are displayed in Figure \ref{fig3},
for the Dirichlet case, and in Figure \ref{fig4} for the Neumann case. For both the Dirichlet and Neumann cases the parameter
$\alpha$ varies in the interval $(\pi/2,\pi)$ and $\beta$ varies, instead, in the interval $\beta\in(0,\pi/2)$.
The curves displayed in bold represent, once again, the values of $\alpha$ and of the position of the piston $a$
for which the Casimir force vanishes. For Dirichlet boundary conditions the Casimir force is positive for values of $(a,\alpha)$ in the region below the
curves in bold, and it is negative for values of $(a,\alpha)$ in the region above them. In the case of Neumann boundary conditions this behavior is reversed.
It is interesting to note that there exist specific values of the parameters $\alpha$ and $\beta$ for which the Casimir force
vanishes for more than one value of the position of the piston $a$. For instance, in the case of Dirichlet boundary conditions
with $\beta\sim 1.32$ and $\alpha\sim 1.8$, Figure \ref{fig3} shows that the Casimir force on the piston vanishes at three points: $a\sim 0.2$, $a\sim0.5$, and $a\sim0.8$.
The first and last of these values are points of unstable equilibrium, while the second one is a point of stable equilibrium. A similar behavior for the
Casimir force on the piston can be observed for Neumann boundary conditions in Figure \ref{fig4}. The existence of more than
one point for which the Casimir force $F_{\textrm{Cas}}(a)$ vanishes is an extremely interesting feature. To the best of our knowledge,
no other piston configuration considered so far has exhibits this type of behavior.

\section{Conclusions}

In this paper we have analyzed the Casimir energy and force for massless scalar fields propagating on a higher dimensional
piston configuration, with geometry $I\times N$, endowed
with general self-adjoint boundary conditions. By exploiting the spectral zeta function regularization method we have obtained explicit
expressions for both the Casimir energy and force. These results are valid for any smooth, compact Riemannian manifold $N$ and for any separated self-adjoint boundary conditions.
The analytic continuation of the spectral zeta function to a neighborhood of the point $s=-1/2$ has been performed as follows: First, the function $\zeta(s)$ is represented in terms of an integral through Cauchy's residue theorem which is valid in a strip of the complex plane. The spectral zeta function is then extended to a domain that
contains the point $s=-1/2$ by adding and subtracting from the integrand a suitable number of asymptotic terms. Since the
manifold $N$ has been kept unspecified throughout the analysis, the final result for the Casimir energy
of the piston is written in terms of the spectral zeta function $\zeta_{N}(s)$ associated with the Laplacian $\Delta_{N}$ on $N$.
In the previous Section we have presented numerical results for the Casimir force on the piston when Dirichlet and
Neumann boundary conditions were imposed at $x=0$ or $x=a$ and general boundary conditions were kept on the remaining ones.

We would like to point out that the result obtained in (\ref{78}) for the Casimir force on the piston is very general
and can easily be used to produce numerical results for special cases other than the Dirichlet and Neumann ones considered in this work.
For instance, one could fix the boundary conditions at the two endpoints of the piston configuration and keep general boundary conditions on
piston itself. Within this setting it could be possible to study the behavior of the Casimir force as the boundary conditions on the
piston vary. In this way one could identify what type of boundary conditions lead to a repulsive or vanishing force.

The results obtained in this work could also be used for studying piston configurations consisting of real materials. In fact, the choice of boundary conditions
is motivated by the need to closely model the physical characteristics of the materials under consideration. The standard boundary conditions, however, might not be suitable
for modeling all materials of interest in applications. By using the method delineated in this work, one can fine-tune the six
independent parameters describing the boundary conditions in order
to model the specific properties of the materials. Once the parameters have been assigned the results (\ref{77}) and (\ref{78}) can be utilized to obtain
and analyze the Casimir energy and force, respectively.

In this paper we have considered a piston configuration with the geometry $I\times N$ with an interval $I$ of the real line and a
smooth compact Reimannian manifold $N$. It would be very instructive to extend the analysis performed here, for instance, to spherical pistons. This would generalize
the results obtained in \cite{dowker11} to include general boundary conditions. The process of analytic continuation of the spectral zeta function
would be similar to the one described in the previous Sections with the exception of the eigenfunctions of the Laplacian on the piston. In the case of a spherical piston
the eigenfunctions would be Bessel functions.
The technique developed here could also be applied to Casimir piston configurations with non-vanishing curvature for which
the geometry of one chamber differs from the geometry of the other (see e.g. \cite{fucci12}).
It would be interesting, in this case, to analyze the behavior of the Casimir force acting on the piston due to the combined effect of general boundary conditions
and difference in the geometry of the chambers.

A very interesting new feature found in some of the examples considered earlier is the presence of
more than one value of the position of the piston $a$ for which the Casimir force vanishes. It would certainly be worthwhile
to further analyze this feature and to determine what other geometries for the piston configuration and boundary conditions lead
to multiple values of the position $a$ where the force is zero.

\begin{acknowledgments}
This research is partially funded by the Research and Creative Activities Committee of East Carolina University
\end{acknowledgments}

\end{document}